\documentclass[aps,prb,twocolumn,floatfix,showpacs,superscriptaddress]{revtex4}

\usepackage{graphicx}
\usepackage{bbm}
\usepackage{amsmath,amssymb}
\newcommand{\be}{\begin{equation}}
\newcommand{\beq}{\begin{equation}}
\newcommand{\ee}{\end{equation}}
\newcommand{\bea}{\begin{eqnarray}}
\newcommand{\eea}{\end{eqnarray}}
\newcommand{\ba}{\begin{array}}
\newcommand{\ea}{\end{array}}

\renewcommand{\vr} {{\bf r}}
\newcommand{\vj} {{\bf j}}
\newcommand{\vs} {{\bf s}}

\begin{document}
\title{Electronic exchange in quantum rings}
\author{E. R{\"a}s{\"a}nen}
\email[Electronic address:\;]{esa.rasanen@jyu.fi}
\affiliation{Institut f{\"u}r Theoretische Physik,
Freie Universit{\"a}t Berlin, Arnimallee 14, D-14195 Berlin, Germany}
\affiliation{Nanoscience Center, Department of Physics,
University of Jyv{\"a}skyl{\"a}, B.O.Box 35, 
FI-40014 Jyv{\"a}skyl{\"a}, Finland}
\affiliation{European Theoretical Spectroscopy Facility (ETSF)}
\author{S. Pittalis}
\email[Electronic address:\;]{pittalis@physik.fu-berlin.de}
\affiliation{Institut f{\"u}r Theoretische Physik,
Freie Universit{\"a}t Berlin, Arnimallee 14, D-14195 Berlin, Germany}
\affiliation{European Theoretical Spectroscopy Facility (ETSF)}
\author{C. R. Proetto}
\affiliation{Institut f{\"u}r Theoretische Physik,
Freie Universit{\"a}t Berlin, Arnimallee 14, D-14195 Berlin, Germany}
\affiliation{European Theoretical Spectroscopy Facility (ETSF)}
%\affiliation{Centro At{\'o}mico Bariloche and Instituto Balseiro, 8400
%S.C. de Bariloche, R{\'i}o Negro, Argentina}
\author{E. K. U. Gross}
\affiliation{Institut f{\"u}r Theoretische Physik,
Freie Universit{\"a}t Berlin, Arnimallee 14, D-14195 Berlin, Germany}
\affiliation{European Theoretical Spectroscopy Facility (ETSF)}

%\date{\today}

\begin{abstract}
Quantum rings can be characterized by a specific radius 
and ring width. For this rich class of physical systems,
an accurate approximation for the exchange-hole potential and
thus for the exchange energy is derived from first principles.
%The description can be viewed as a density functional within the
%current-dependent meta-generalized-gradient approximation 
%specialized for the ring topology. 
Excellent agreement with the
exact-exchange results is obtained regardless of the ring parameters, 
total spin, current, or the external magnetic field.
The description can be applied as a density functional
outperforming the commonly used local-spin-density 
approximation, which is here explicitly shown to break 
down in the quasi-one-dimensional limit. 
The dimensional crossover, which is of extraordinary importance
in low-dimensional systems, is fully captured by our functional.
%In contrast with the exact-exchange approach in
%the Kohn-Sham scheme, our functional 
%forms a practical basis for the development of
%corresponding correlation-energy functionals.
\end{abstract}


\maketitle

%\input{qdxholecurvature}

%\section{Introduction}

Ring-shaped quantum systems such as semiconductor
quantum rings (QRs) have attracted wide interest
both theoretically and experimentally.
Presently, QRs can be fabricated using
a variety of techniques.~\cite{lorke,fuhrer,keyser}
The tunability of size, shape, and electron 
number in QRs suggests applications in 
the field of quantum-information technology.
In particular, one can exploit the 
well-known Aharonov-Bohm effect~\cite{ab} by 
external magnetic fields,~\cite{interferometer} 
or control the electronic states by short laser
pulses within the decoherence time.~\cite{control}

Many-electron QRs have been studied theoretically using 
various approaches, e.g., model Hamiltonians,~\cite{beginners,fogler} 
exact diagonalization,~\cite{niemela,sigga} 
quantum Monte Carlo,~\cite{sigga,emperador_DMC}
and density-functional theory~\cite{emperador_LSDA,aichinger,aichinger2} (DFT).
Within spin-DFT (SDFT) or current-SDFT applied to QRs, the 
exchange and correlation energies and potentials are commonly
calculated using the two-dimensional (2D) local 
spin-density approximation~\cite{attaccalite} (LSDA).
%with the
%exchange functional by Rajagopal and Kimball~\cite{rajagopal} 
%combined with the corresponding correlation functional 
%parametrized using Monte Carlo methods~\cite{attaccalite}.
For relatively wide QRs, LSDA performs reasonably
well,~\cite{emperador_LSDA,aichinger} which is not surprising
in view of the good performance of the LSDA in the case
of 2D quantum dots.~\cite{qd} However, in the 
quasi-one-dimensional limit the 2D-LSDA is expected to
fail similarly to the breakdown of the
three-dimensional LSDA in the quasi-2D limit.~\cite{kim_pollack}
Hence, in order to benefit from the efficiency 
of DFT methods in various QRs,
accurate density functionals for exchange and correlation are needed. 
Methods based on exact-exchange (EXX) functionals seem an attractive 
possibility.~\cite{nicole} 
%but currently they are in lack of 
%practical ways to include the correlation effects.
However, most computational schemes exploting EXX for finite systems
suffer from numerical problems,~\cite{nicole,EXX_problems} 
which, ultimately, prevents the method from being applied 
to large electron numbers.

In this paper we derive an accurate
method to calculate the exchange-hole potential and the exchange 
energy of QRs. The resulting density functional is simpler than
EXX, but yet compatible with correlation functionals based on
the modeling of the correlation hole.~\cite{corr_paper}
Our derivation follows
the strategy originally proposed for atoms by Becke and Roussel,~\cite{becke}
in which the averaged exchange hole of a suitable single-electron 
wavefunction is adapted to a general $N$-electron system by
examining the short-range behavior of the exchange hole.
%Recently, this approach has been used to develop
%exchange functionals for finite 2D systems~\cite{expaper}.
%In this respect, the derivation here can be seen as an extension 
%to 2D systems having a ring topology.
%The resulting set of formulae can be applied in the Kohn-Sham (KS)
%scheme of DFT within the current-dependent meta-generalized-gradient approximation 
%(meta-GGA), which is adapted to the 2D ring topology. 
Recently, a similar approach has been used to develop exchange
functionals for finite 2D systems,~\cite{expaper} which 
can be reproduced as a special case of this work.
In the numerical examples we demonstrate the
accuracy of the functional against exact-exchange results 
and underline the considerable improvement over the LSDA, 
particularly in the quasi-1D limit. 

In the framework of SDFT
%, the total energy can be written as a 
%functional of the
%spin densities $\rho_{\uparrow}(\vr)$ and $\rho_{\downarrow}(\vr)$ as
%\begin{eqnarray}
%E_{\rm tot}[\rho_{\uparrow},\rho_{\downarrow}] & = & T_s[\rho_{\uparrow},\rho_{\downarrow}] + 
%E_{\rm H}[\rho] + E_{xc}[\rho_{\uparrow},\rho_{\downarrow}] \nonumber \\
%& + & \sum_{\sigma=\uparrow,\downarrow}\int{d\vr} \; v_{\sigma}(\vr)
%\rho_{\sigma}(\vr), 
%\label{etot}
%\end{eqnarray}
%where $T_s[\rho_{\uparrow},\rho_{\downarrow}]$ is the KS 
%kinetic-energy functional, $v_{\sigma}(\vr)$ is the external (local)
%scalar 
%potential acting upon the interacting system, $E_{\rm H}[\rho]$ 
%is the Hartree energy of the total charge density 
%$\rho(\vr)=\rho_{\uparrow}(\vr)+\rho_{\downarrow}(\vr)$, and  
%$E_{xc}[\rho_{\uparrow},\rho_{\downarrow}]= E_{x}[\rho_{\uparrow},\rho_{\downarrow}] + E_{c}[\rho_{\uparrow},\rho_{\downarrow}]$ is the exchange-correlation 
%energy functional. 
the exact exchange energy functional of the spin
densities $\rho_{\uparrow}(\vr)$ and $\rho_{\downarrow}(\vr)$ can be
written in (effective) atomic units~\footnote{
For the effective atomic units $a^*_0 =(\epsilon/m^*)a_0$ and
${\rm Ha}^*  = (m^*/\epsilon^2)\,{\rm Ha}$, we use the material
parameters of GaAs, $m^*=0.067\,m_e$ and $\epsilon=12.4\,\epsilon_0$.
All the formulas and results presented here are the same in conventional
Hartree atomic units -- apart from Fig.~\ref{nlarge}(a) requiring
different scaling with respect to the magnetic field.}
(a.u.) as
\begin{equation}
E_x [\rho_{\uparrow},\rho_{\downarrow}] = - \frac{1}{2} \sum_{\sigma=\uparrow,\downarrow} \int d\vr_1 \int d\vr_2 \frac{\rho_{\sigma}(\vr_1)}{|\vr_1-\vr_2|}
h^{\sigma}_{x}(\vr_1,\vr_2),
\label{EX_2}
\end{equation}
where
\begin{equation}
h^{\sigma}_{x}(\vr_1,\vr_2) =
\frac{|\sum_{k=1}^{N_\sigma}\psi^*_{k,\sigma}(\vr_1)\psi_{k,\sigma}(\vr_2)|^2}
{\rho_{\sigma}(\vr_1)}
\label{xhole}
\end{equation}
is the exchange-hole function. Here we assume that 
the noninteracting ground state is nondegenerate and hence takes the 
form of a single Slater determinant constructed from
the KS orbitals, $\psi_{k,\sigma}$. 

%In order to construct a density functional for the exchange energy 
%in a QR, 
Next we look for an approximation for the {\em cylindrical} 
average of the exchange hole. It is defined around $\vr_1$ 
with respect to $\vs=\vr_2-\vr_1$ as~\cite{expaper}
\begin{eqnarray}
\bar{h}^{\sigma}_x(\vr_1,s) & = & \frac{1}{2\pi}\int_{0}^{2\pi}
d\phi_s \,h^{\sigma}_x(\vr_1,\vr_1+\vs).
%\nonumber \\
%& = & \frac{1}{2\pi}\int_{0}^{2\pi} d\phi_s\, e^{\vs\cdot\nabla'} h^\sigma_x(\vr_1,\vr')|_{\vr'=\vr_1}.
\label{hbar}
\end{eqnarray}
In the following, we compute ${\bar h}^{\sigma}_{x}$ {\em exactly} for 
a single-electron wavefunction of a QR. We consider a
2D QR defined by a radial confining potential of the form~\cite{tan}
\begin{equation}
%V(r)=a_1/r^2+a_2 r^2 - 2\sqrt{a_1 a_2}.
V(r)=\frac{M^2}{2 r^2} + \frac{\alpha^4}{2}r^2 - M\alpha^2,
\label{conf}
\end{equation}
where $M\geq 0$ and $\alpha>0$ are constants. 
Note that $M=0$ corresponds to a harmonic quantum dot.~\cite{qd}
Regarding QR fabrication,~\cite{lorke,fuhrer,keyser} the confinement 
given above is realistic: On one hand, 
the electrons cannot enter the center area
described by the strongly peaked ``antidot'' [first term in $V(r)$], 
and, on the other hand,
the edge of the QR is described by a soft, parabolic
confinement [second term in $V(r)$]. In this tunable model, 
both the radius and width of the QR can be changed 
independently by varying $M$ and $\alpha$ (see below).
%Figure~\ref{width}
%\begin{figure}
%\includegraphics[width=0.9\columnwidth]{width_radius.eps}
%\caption{(color online) Relation between the parameter set
%($\alpha$, $M$) in the external confining potential [Eq.~(\ref{conf})]
%and the ring radius $r_0$ and width $\Delta r$ (see the inset).
%The curves correspond to fixed values of $M$.
%}
%\label{width}
%\end{figure}
%shows explicitly the relation between ($r_0,\Delta r$)
%and ($M,\alpha$).

\begin{figure}
\includegraphics[width=0.8\columnwidth]{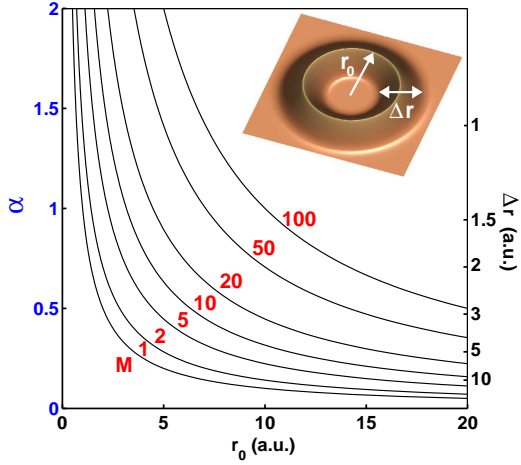}
\caption{(color online) Relation between the parameter set
($\alpha$, $M$) in the external confining potential [Eq.~(\ref{conf})]
and the single-electron ring radius $r_0$ and width $\Delta r$ (see the inset).
The curves correspond to fixed values of $M$.
}
\label{width}
\end{figure}

The eigenfunctions and eigenvalues for a single electron confined by 
$V(r)$ in Eq.~(\ref{conf}) can be solved analytically.~\cite{tan}
Setting the radial and angular quantum numbers to zero, $n=m=0$,
yields a wavefunction
\begin{equation}
\psi_{\sigma}(\vr)=\frac{\alpha^{M+1}}{\sqrt{\pi M!}}\;r^{M}\, e^{-\alpha^2 r^2/2}.
\label{wf}
\end{equation}
%where $M=\sqrt{2 a_1}$ and $\alpha=(2 a_2)^{1/4}$.
We emphasize that this expression, normalized for each $\alpha$,
is the {\em general}
single-electron ground-state wavefunction 
of a QR having a potential minimum at
$r=r_0 = \sqrt{M}/\alpha$, which corresponds to the
ring radius.
The width of the single-electron QR having the energy 
$E_{0}=\alpha^2$ can be approximated by 
$\Delta r = \sqrt{2}/\alpha$. 
 %Second, the inner and outer radius of the single-electron QR, 
%$$r_{-}$ and $r_{+}$, 
%can be approximated by~\cite{tan}
%\begin{equation}
%r_{\pm} = \alpha^{-2}\left(M\alpha^2+E_F \pm \sqrt{2 E_F M \alpha^2 +
%  E_F^2}\right)^{1/2},
%r_{\pm} = \left(M+1 \pm \sqrt{2M+1}\right)^{1/2},
%\label{width}
%\end{equation}
%so that the width of the QR is $\Delta r = r_{+}-r_{-}$. 
%In the general $N$-electron case, the ring width can be expressed 
%in terms of the Fermi energy, see Eq. (4) in Ref.~\cite{tan}.
In Fig.~\ref{width}
we show explicitly the relation between the {\em single-electron}
radius and width of the QR ($r_0,\Delta r$) and the parameters
($M,\alpha$) in the external confining potential given in Eq.~(\ref{conf}).

The exact exchange-hole function for the {\em single-electron} wavefunction
in Eq.~(\ref{wf}) becomes
\begin{eqnarray}
h^{\sigma}_{x,1}(\vr_1,\vr_2) & = & 
\psi^*_{\sigma}(\vr_2)\psi_{\sigma}(\vr_2)= 
\rho_\sigma(\vr_2) \nonumber \\
& = & \frac{\alpha^{2(M+1)}}{\pi M!}\;r_2^{2M}\, e^{-\alpha^2 r_2^2}.
\end{eqnarray}
To calculate the cylindrical average as defined in Eq.~(\ref{hbar}), 
we first set $\vr_1=\vr$ and $\vr_2=\vr+\vs$. As a result we get
%\begin{eqnarray}
%\bar{h}^\sigma_{x,1}(\vr,s) & = & 
% \frac{1}{2\pi} \int_{0}^{2\pi} d\phi_s \,\rho_\sigma(\vr+\vs) \nonumber \\
%& = & 
%\frac{\alpha^{2(M+1)}}{\pi M!}e^{-\alpha^2(r^2+s^2)} (r^2+s^2)^M
%\nonumber \\
%& \times & \sum_{k=0}^M (-1)^k \binom{M}{k}\left(\frac{2 r
%    s}{r^2+s^2}\right)^k \nonumber \\ 
%& \times & \frac{d^k}{d z^k} I_0(z){\Big|}_{z=2\alpha^2 r s},
%\label{hbar1}
%\end{eqnarray}
\begin{multline}\label{hbar1}
\bar{h}^\sigma_{x,1}(\vr,s) =
 \frac{1}{2\pi} \int_{0}^{2\pi} d\phi_s \,\rho_\sigma(\vr+\vs) \\
= \frac{\alpha^{2(M+1)}}{\pi M!}e^{-\alpha^2(r^2+s^2)} (r^2+s^2)^M \\
 \times  \sum_{k=0}^M (-1)^k \binom{M}{k}\left(\frac{2 r
    s}{r^2+s^2}\right)^k \frac{d^k}{d z^k} I_0(z){\Big|}_{z=2\alpha^2 r s},
\end{multline}
where $I_0(z)$ is the zeroth-order modified Bessel 
function of the first kind, and $M$ is now an integer.
%Its derivatives can be 
%computed conveniently using the identity~\cite{handbook}
%\begin{eqnarray}
%\frac{d^k}{d z^k}
%I_0(z) & = & \frac{1}{2^k}\Bigg[\binom{k}{0}I_{-k}(z)+\binom{k}{1}I_{-k+2}(z) \nonumber \\
%& + & \binom{k}{2}I_{-k+4}(z)+\ldots+\binom{k}{k}I_{k}(z)  \Bigg]
%\label{bessel}
%\end{eqnarray}

Next, we apply $\bar{h}^\sigma_{x,1}$ as a {\em general} model
for the averaged exchange hole of an {$N$-electron} 
system.~\cite{becke,expaper}
The model is required to locally reproduce the short-range 
behavior of the exchange hole. Therefore, we parametrize $\alpha^2$
and $r^2$ into functions of $r$ by setting 
$\alpha^2\rightarrow a(r)$ and $r^2\rightarrow b(r)$.
Equation~(\ref{hbar1}) can be now rewritten as
\begin{multline}\label{hbarr}
\bar{h}^\sigma_{x}(a,b;s) = 
\frac{a^{M^*+1}}{\pi {M^*}!}e^{-a(b+s^2)} (b+s^2)^{M^*} \\
\times \sum_{k=0}^{M^*} (-1)^k \binom{M^*}{k}\left(\frac{2 \sqrt{b}
    s}{b+s^2}\right)^k\frac{d^k}{d z^k} I_0(z){\Big|}_{z=2a\sqrt{b} s}.
\end{multline}
%\begin{eqnarray}
%\bar{h}^\sigma_{x}(a,b;s) & = & 
%\frac{a^{M^*+1}}{\pi {M^*}!}e^{-a(b+s^2)} (b+s^2)^{M^*}
%\nonumber \\
%& \times & \sum_{k=0}^{M^*} (-1)^k \binom{M^*}{k}\left(\frac{2 \sqrt{b}
%    s}{b+s^2}\right)^k \nonumber \\ 
%& \times & \frac{d^k}{d z^k} I_0(z){\Big|}_{z=2a\sqrt{b} s}.
%\label{hbarr}
%\end{eqnarray}
%where we have omitted the lower index ``$1$''. 
where $M^*=2(\Delta r^*/r^*_0)^{-2}$ is the nearest integer describing the
characteristic ratio between the effective width and the radius 
of the $N$-electron QR. As we will show below, $r_0^*$ and $\Delta r^*$
can be extracted from the spin density. As a consequence, $M^*$
can be considered as a (spin) density functional $M^*[\rho_\sigma]$.
The short-range behavior with respect to $s$ can be obtained
from the first two non-vanishing terms in the Taylor expansion 
in Eq.~(\ref{hbar}):
\begin{equation}
\bar{h}^{\sigma}_x(\vr,s) = \rho_{\sigma}(\vr) + s^2 C^{\sigma}_{x}(\vr) + \ldots ,
% & = & \rho_{\sigma}(\vr)
%+ \frac{s^2}{4} \nabla'^2 h^{\sigma}_x(\vr,\vr')|_{\vr'=\vr} +
%\ldots \nonumber \\
%& = & \rho_{\sigma}(\vr) + s^2 C^{\sigma}_{x}(\vr) + \ldots ,
\label{taylor2}
\end{equation}
where 
\begin{equation}
C^{\sigma}_x = \frac{1}{4}\left[ \nabla^2 \rho_{\sigma} -2\tau_\sigma
+ \frac{1}{2}\frac{\left( \nabla \rho_\sigma \right)^2}{\rho_\sigma}
+ 2 \frac{\vj^2_{p,\sigma}}{\rho_\sigma} \right],
\label{C}
\end{equation}
is the local curvature of the exchange
hole in 2D~\cite{expaper} around the given reference 
point $\vr$ (argument omitted). Here the (double of the) kinetic-energy density,
%\begin{equation}
$\tau_\sigma=\sum_{k=1}^{N_\sigma} |\nabla\psi_{k,\sigma}|^2$,
%\end{equation}
and the spin-dependent paramagnetic current density,
%\begin{equation}
$\vj_{p,\sigma}=\frac{1}{2i}\sum_{k=1}^{N_\sigma} \left[
  \psi^*_{k,\sigma} \left(\nabla \psi_{k,\sigma}\right) - \left(\nabla \psi^*_{k,\sigma}\right) \psi_{k,\sigma} \right]$,
%\end{equation}
depend explicitly on the KS orbitals, and hence implicitly
on the spin densities. Defining $y(r):=a(r)b(r)$, the
zeroth-order term in Eq.~(\ref{taylor2}) yields %(argument $r$ omitted below)
\begin{equation}
\rho_{\sigma}=\frac{a}{\pi {M^*}!} y^{M^*} e^{-y},
\label{zero}
\end{equation}
and the second-order term gives
\begin{equation}
C_x^{\sigma}=\frac{a^2}{\pi M^*!}y^{M^*-1}\big[(y-M^*)^2-y\big] e^{-y}.
\label{second}
\end{equation}
Combining Eqs.~(\ref{zero}) and (\ref{second}) leads to
\begin{equation}
M^*!\,y^{-(M^*+1)}\big[(y-M^*)^2-y\big] e^{y}=\frac{C_x^{\sigma}}{\pi\rho^2_{\sigma}},
\label{y}
\end{equation}
from which $y$ can be solved numerically. 
%Solution of $y$ yields $a$
%through Eq.~(\ref{zero}), and $b$ follows from $b=y/a$.
%Hence, we can compute the averaged exchange hole from
Now, we can compute $a$ and $b$ and thus
the averaged exchange hole from Eq.~(\ref{hbarr}). 
The exchange-hole potential is given by
\begin{equation}
U^{\sigma}_x(\vr)= - 2 \pi \int_{0}^{\infty}  ds \,\bar{h}^{\sigma}_x(a,b;s),
\label{xenergydensity}
\end{equation}
from which the exchange energy can be calculated as
\begin{equation}
E_x[\rho_{\uparrow},\rho_{\downarrow}] =  \frac{1}{2} \sum_{\sigma} \int d\vr \,\rho_{\sigma}(\vr) 
U^{\sigma}_x(\vr).
\label{xenergy}
\end{equation}
From Eqs.~(\ref{hbarr}) and (\ref{xenergydensity}) it can be shown 
that our calculation scheme preserves also the exact long-range
behavior: $U_x^{\sigma}(r\rightarrow\infty)\rightarrow -1/r $.~\cite{new}

%We point out that
%as the only ingredients in the above approximation
%we have (i) a set of KS orbitals $\psi_{k,\sigma}$,
%and (ii) a value for $M^*$, which couples with the
%characteristic ratio between the QR width and radius
%as $\Delta r/r_0=\sqrt{2/M^*}$.
%On the whole, the above {\em ab initio} derivation has led to a 
%current-dependent meta-GGA functional. 
To summarize our method, the calculation of the 
exchange energy of an $N$-electron QR consists of the following steps:
\begin{itemize}
\setlength{\labelwidth}{5truecm}
\item[(1)] Use the KS orbitals $\psi_{k,\sigma}$ to
calculate the spin densities $\rho_{\sigma}$, 
the local curvature of the exchange hole $C^{\sigma}_x$ from
Eq.~(\ref{C}), and to estimate $M^*$.
%\item[(2)] Estimate $M^*$ from $\rho_\sigma$ (see below).
\item[(2)] Compute $y(r)$ numerically from Eq.~(\ref{y}).
\item[(3)] Calculate $a(r)$ from Eq.~(\ref{zero}) and $b(r)$ from 
the relation $b(r)=y(r)/a(r)$.
\item[(4)] Calculate the averaged exchange-hole function 
from Eq.~(\ref{hbarr}).
\item[(5)] Finally calculate the exchange-hole potential and the
  exchange energy from Eqs.~(\ref{xenergydensity}) and
  (\ref{xenergy}), respectively.
\end{itemize}
%On the whole, the above {\em ab initio} derivation has led to a 
%current-dependent meta-GGA functional.
%An value for $M^*[\rho]$ is obtained by 

%designed to perform particularly well
%for the full class of QRs. We emphasize that the functional
%was derived {\em ab initio} without external fitting parameters 
%all the required information is incorporated in $\Delta r/r_0$ yielding
%a value for $M*$.

Next we test the performance of our functional
in a few examples. As the reference results we use the
exchange-hole potentials and exchange energies of the EXX
calculations within the rather
accurate Krieger-Li-Iafrate (KLI) approximation.~\cite{KLI}
We compare the results also to the 2D-LSDA. 
%where the 
%exchange energy per particle~\cite{rajagopal,attaccalite}
%%$\epsilon^{\sigma}_x\propto \rho_{\sigma}^{1/2}$,
%can be directly associated with $U^{\sigma}_x$ as the integrand 
%of $E_x$. 
Both the EXX and LSDA
results are obtained using the {\tt octopus} code.~\cite{octopus}
The converged EXX orbitals are used 
as the input KS orbitals in our functional (and in the LSDA).
Alternatively, also the LSDA orbitals can be used as input,
which in most cases leads to only minor changes in the results. 

\begin{figure}
\includegraphics[width=0.6\columnwidth]{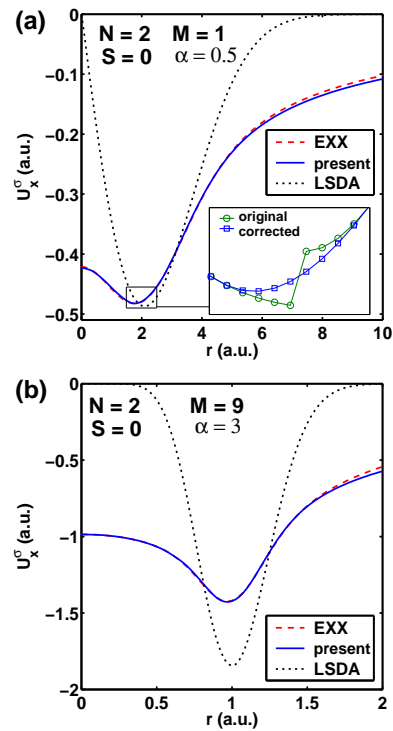}
\caption{(color online) Exchange-hole potentials of two-electron quantum rings
(total spin $S=0$) defined by parameters (a) $M=1$, $\alpha=0.5$,
and (b) $M=9$, $\alpha=3$. The dashed line shows the exact-exchange (EXX)
result, the solid line is the result of the present work, and the dotted
line corresponds to the exchange-energy density in the local 
spin-density approximation (LSDA).
}
\label{n2}
\end{figure}

\begin{figure}
\includegraphics[width=0.6\columnwidth]{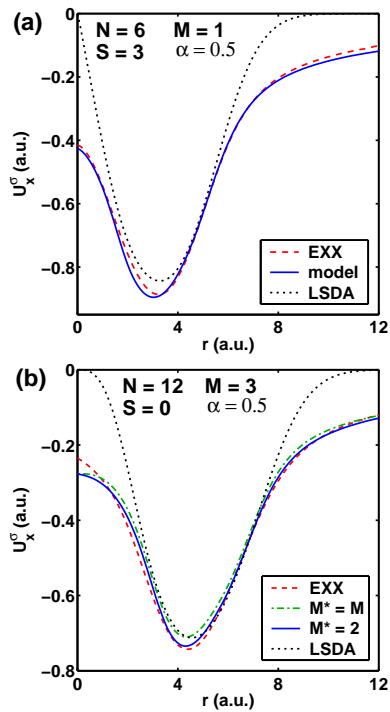}
\caption{
(color online) Same as in Fig.~\ref{n2} but for (a) a fully spin-polarized 
current-carrying six-electron quantum ring at $B=6$ T and for (b) 
a zero-current 12-electron quantum ring.
}
\label{nlarge}
\end{figure}

Figure~\ref{n2} 
shows the exchange-hole potentials for 
two-electron singlet states of two QRs defined
by (a) ($M,\alpha$)=($1,0.5$) and (b) ($M,\alpha$)=($9,3$), respectively.
Note that here we have set $M^*=M$ as the first
approximation.
In both cases we find excellent agreement between the EXX 
(dashed lines) and the present work (solid lines). Both the 
large-$r$ limit, where $U^{\sigma}_x$ decays 
as $-1/r$, and also the $r\rightarrow 0$ limit, are
correctly reproduced. 

The exchange energies per particle of the LSDA 
(dotted lines in Fig.~\ref{n2}), which are directly
comparable to $U_x^\sigma$, deviate significantly
from the EXX and from our approximation. The
inability of the LSDA to yield the correct shape of the curve
is due to the simple density-dependent expression of the
exchange energy per particle in the LSDA, 
i.e., $\epsilon^{\rm LSDA}_{x,\sigma}\propto \rho_{\sigma}^{1/2}$,
leading to differences also in the exchange energy.
In the case of Fig.~\ref{n2}(a), we find $E^{\rm LSDA}_x=-0.389$,
whereas $E^{\rm EXX}_x=-0.409$ and $E^{\rm present}_x=-0.408$.
When the ratio $\Delta r/r_0$ is decreased to one third,
i.e., $M=1\rightarrow 9$, the deviation of 
the LSDA becomes more pronounced as shown in Fig.~\ref{n2}(b). 
Now we find $E^{\rm LSDA}_x=-1.502$ vs. $E^{\rm EXX}_x=E^{\rm present}_x=-1.300$.
In other words, the above change in $M$ increases
the relative error of the LSDA exchange energy from $5\%$ to $16\%$.
Hence, these results demonstrate the 
breakdown of the 2D-LSDA in the quasi-one-dimensional limit.
%The inability of the LSDA to yield the correct shape
%of $U^{\sigma}_x$ leads also to erroneous exchange energies:

% M = 1:
% E_x_EXX  = -0.4089
% E_x_here = -0.4082
% E_x_LSDA = -0.3885

% M = 9
% E_x_EXX  = -1.2996
% E_x_here = -1.2995
% E_x_LSDA = -1.5022

At this point we make two remarks of the numerical procedure. 
First, Eq.~(\ref{y}) has two solutions for $y(r)$, from which
we choose the smaller one for $r<r[\rho_{\rm max}]$
and the larger one otherwise. Since these
two solutions do not generally coincide at 
$r=r[\rho_{\rm max}]$, we extrapolate $U_x^{\sigma}(r)$ around
this point as visualized in the inset of Fig.~\ref{n2}.
This smoothing procedure is done such that $E_x$ is not altered. 
%and it could be easily automatized in the self-consistent 
%applications. 
Second, the application of the overall scheme
might be numerically cumbersome in the small-$r$ regime
when $M$ (and/or $M^*$) is large and the density is very small.
To overcome this problem, we take the advantage of the 
exact property of our model,
%\be
$U_x^{\sigma}(r\rightarrow 0)=-(2M)!\,(M!)^{-2}\,2^{-2M}\,\alpha\sqrt{\pi}+O(r^2)$,
%\ee
which is valid for a general $N$-electron case with $M>1$.~\cite{new}

To demonstrate the generality of our method,
we show in Fig.~\ref{nlarge}
the exchange-hole potentials for (a) a spin-polarized 
current-carrying six-electron QR at $B=6$ T and for (b) a 
12-electron QR.
Similarly to the previous examples, we find excellent
agreement with the EXX. 
The six-electron case yields $E^{\rm EXX}_x=-2.226$, 
$E^{\rm present}_x=-2.238$, and $E^{\rm LSDA}_x=-2.114$.
In the case of 12 electrons, we plot results for both
$M^*=M=3$ and for $M^*[\rho]=2\approx 2(\Delta r^*/r^*_0)^{-2}$,
where the effective radius $r^*_0$ corresponds to the
point where the cumulative density reaches $50\,\%$,
and the effective width $\Delta r^*$, centered at $r^*_0$, 
covers $90\,\%$ of the total density. We find that in
the latter case the agreement with the EXX is better.
%justifying the use of the parameter $M^*$ as a density
%functional.
%For the large QR in Fig.~\ref{nlarge}(b) 
%a reliable EXX result is not available due to 
%well-known but, so far, not completely resolved
%numerical instabilities~\cite{nicole,asym}.
%Nevertheless, our functional performs well in the full range
%of $r$, and is assumed to provide reliable reference data
%for large QRs which are beyond the reach of EXX. 
%On the other hand, the LSDA result for the large QR of 
%Fig.~\ref{nlarge}(b) is, as expected, close to our 
%functional, except for small and large $r$.

To conclude, we have derived, from first principles, an accurate 
and general approximation
%within the current-dependent meta-generalized-gradient
%approximation 
for the exchange-hole potential and hence for the exchange energy 
in quantum rings.
Excellent agreement with the exact-exchange results is obtained
regardless of the 
ring geometry, number of electrons, spin polarization, and currents.
Moreover, we have demonstrated that, in contrast to the local-density
approximation, our functional can deal with the physically
relevant dimensional
crossover between two and one dimensions.
%Moreover, the 
%functional provides reliable reference data for large 
%quantum rings which are beyond reach of the present exact-exchange
%methods. 
Our approach is suitable for the development
of correlation functionals
%, e.g.,
%in the framework of Becke~\cite{becke_corr},
by considering the exact properties of the corresponding 
correlation-hole potentials.~\cite{corr_paper}

% E_x_EXX  = -2.2259
% E_x_here = -2.2383
% E_x_LSDA = -2.1135

\begin{acknowledgments}
This work was supported by the Deutsche 
Forschungsgemeinschaft, the EU's Sixth Framework
Programme through the Nanoquanta Network of
Excellence (NMP4-CT-2004-500198), and the
Academy of Finland.
C. R. P. was supported by the European
Community through a Marie Curie IIF (MIF1-CT-2006-040222) 
and CONICET of Argentina through PIP 5254.
\end{acknowledgments}


\begin{thebibliography}{ll}

\bibitem{lorke}
A. Lorke, R.~J. Luyken, A.~O. Govorov, J.~P. Kotthaus, J.~M. Garcia, and P.~M.
Petroff, Phys. Rev. Lett. {\bf 84},  2223  (2000).

\bibitem{fuhrer}
A. Fuhrer, S. L{\"u}scher, T. Ihn, T. Heinzel, K. Ensslin, W. Wegscheider, and
M. Bichler, Nature (London) {\bf 413},  822  (2001).

\bibitem{keyser}
U.~F. Keyser, C. F{\"u}hner, S. Borck, R.~J. Haug, M. Bichler, G. Abstreiter,
and W. Wegscheider, Phys. Rev. Lett. {\bf 90},  196601  (2003).

\bibitem{ab} Y. Aharonov and D. Bohm, Phys. Rev. {\bf 115}, 485 (1959).

\bibitem{interferometer} See, e.g., M. Sigrist, A. Fuhrer, T. Ihn, 
K. Ensslin, S. E. Ulloa, W. Wegscheider, and M. Bichler,
Phys. Rev. Lett. {\bf 93}, 066802 (2004).

\bibitem{control} E. R{\"a}s{\"a}nen, A. Castro, 
J. Werschnik, A. Rubio, and E. K. U. Gross,
Phys. Rev. Lett. {\bf 98}, 157404 (2007).

\bibitem{beginners} S. Viefers, P. Koskinen, P. Singha Deo, and
M. Manninen, Physica E (Amsterdam) {\bf 21}, 1 (2004).

\bibitem{fogler} M. M. Fogler and E. Pivovarov, 
Phys. Rev. B {\bf 72}, 195344 (2005).

\bibitem{niemela}
K. Niemel{\"a}, P. Pietil{\"a}inen, P. Hyv{\"o}nen, and T. Chakraborty,
  Europhys. Lett. {\bf 36},  533  (1996).

\bibitem{sigga} S. S. Gylfadottir, A. Harju, T. Jouttenus, C. Webb, 
New J. of Physics {\bf 8}, 211 (2006).

\bibitem{emperador_DMC}
A. Emperador, F. Pederiva, and E. Lipparini, Phys. Rev. B {\bf 68},  115312
  (2003).

\bibitem{emperador_LSDA}
A. Emperador, M. Pi, M. Barranco, and E. Lipparini, 
Phys. Rev. B {\bf 64},  155304  (2001).

\bibitem{aichinger} M. Aichinger, S. A. Chin, 
E. Krotscheck, and E. R{\"a}s{\"a}nen, 
Phys. Rev. B {\bf 73}, 195310 (2006).

\bibitem{aichinger2} E. R\"as\"anen and M. Aichinger,
J. Phys.: Cond. Matt. {\bf 21}, 025301 (2009).

\bibitem{attaccalite}
C. Attaccalite, S. Moroni, P. Gori-Giorgi, and G. B. Bachelet, 
Phys. Rev. Lett. {\bf 88}, 256601 (2002).

%\bibitem{dft} For a review, see, e.g., R. M. Dreizler and E. K. U. Gross,
%{\it Density functional theory} (Springer, Berlin, 1990).

%\bibitem{functionals} For a review, see, J. P. Perdew and S. Kurth,
%in {\it A Primer in Density Functional Theory} (Springer, Berlin, 2003).

\bibitem{qd}
  For a review, see, e.g., L. P. Kouwenhoven, D. G. Austing, and S. Tarucha, Rep.
  Prog. Phys. {\bf 64}, 701 (2001); S. M. Reimann and M. Manninen, Rev. Mod.
  Phys. {\bf 74}, 1283 (2002).

\bibitem{kim_pollack} 
Y.-H. Kim, I.-H. Lee, S. Nagaraja, J.-P. Leburton, R. Q. Hood,
and R. M. Martin, Phys. Rev. B {\bf 61}, 5202 (2000); L. Pollack 
and J. P. Perdew, J. Phys.: Condens. Matter {\bf 12}, 1239 (2000).

\bibitem{nicole}
N. Helbig, S. Kurth, S. Pittalis, E. R\"as\"anen, and E. K. U. Gross,
Phys. Rev. B {\bf 77}, 245106 (2008).

\bibitem{EXX_problems} 
See, e.g., 
T. Heaton-Burgess, F. A. Bulat, and W. Yang, Phys. Rev. Lett. {\bf 98},
256401 (2007); V. N. Staroverov and G. E. Scuseria, 
J. Chem. Phys. {\bf 124}, 141103 (2006); {\em ibid} {\bf 125}, 
081104 (2006); D. R. Rohr, O. V. Gritsenko, and E. J. Baerends, 
J. Mol. Structure: THEOCHEM {\bf 72}, 762 (2006);
A. Hesselmann, A. W. G\"otz, F. Della Sala, and A. G\"orling,
J. Chem. Phys. {\bf 127}, 054102 (2007).

\bibitem{corr_paper} S. Pittalis, E. R\"as\"anen, C. Proetto, 
and E. K. U. Gross, Phys. Rev. B (in print).

\bibitem{becke} 
A. D. Becke and M. R. Roussel, Phys. Rev. A {\bf 39}, 3761 (1989).

\bibitem{expaper} S. Pittalis, E. R{\"a}s{\"a}nen, 
N. Helbig, and E. K. U. Gross, Phys. Rev. B {\bf 76}, 235314 (2007).

\bibitem{tan} W.-C. Tan and J. Inkson, Semicond. Sci. Technol. {\bf
    11}, 1635 (1996).

%\bibitem{becke2} 
%A. D. Becke, Int. J. Quant. Chem. {\bf 23}, 1915 (1983).

%\bibitem{Dobson:93} 
%J. F. Dobson, J.~Chem. Phys. {\bf 98}, 8870 (1993).

\bibitem{new} S. Pittalis {\em et al.} (unpublished).

\bibitem{KLI} J. B. Krieger, Y. Li, and G. J. Iafrate, Phys. Rev. A
  {\bf 46}, 5453 (1992).

\bibitem{rajagopal}
A. K. Rajagopal and J. C. Kimball,
Phys. Rev. B {\bf 15}, 2819 (1977).

\bibitem{octopus} A. Castro,
H. Appel, M. Oliveira, C. A. Rozzi, X. Andrade,
F. Lorenzen, M. A. L. Marques, E. K. U. Gross, and A. Rubio,
Phys. Stat. Sol. (b) {\bf 243}, 2465 (2006).

%\bibitem{asym} See, e.g.,
%T.~H.~Burgess, F.~A. Bulat, and
%W. Yang, Phys. Rev. Lett. {\bf 98}, 256401 (2007);
%V.~N.~Staroverov and G.~E.~Scuseria, J. Chem. Phys. {\bf 124}, 141103
%(2006); {\em ibid} {\bf 125}, 081104 (2006); D.~R.~Rohr {\em et al.},
%THEOCHEM {\bf 72}, 762 (2006); A. Hesselmann {\em et al.},
%J. Chem. Phys. {\bf 127}, 054102 (2007).

%\bibitem{becke_corr} A. D. Becke, J. Chem. Phys. {\bf 88}, 1053 (1988).

\end{thebibliography}
\end{document}